\documentstyle[12pt,aasms4]{article}

\begin{document}
\newcommand{\gray}{$\gamma$-ray\ }
\newcommand{\grays}{$\gamma$-rays\ }
\newcommand{\etal}{{\it et al.}}
\newcommand{\lya}{Ly$\alpha$\ }

\title{Absorption of High Energy Gamma Rays by Interactions With
Starlight Photons in Extragalactic Space at High Redshifts and
the High Energy Gamma-Ray Background}

\author{M.H. Salamon}
\affil{Physics Department, University of Utah}
\authoraddr{Salt Lake City, UT 84112}
\and
\author{F.W. Stecker}
\affil{Laboratory for High Energy Astrophysics,
NASA/Goddard Space Flight Center}
\authoraddr{Greenbelt, MD 20771}

\begin{abstract}

In this paper, we extend previous work on the absorption of high energy
\grays in intergalactic space by calculating the absorption of 10 to
500 GeV \grays at high redshifts. This calculation requires
the determination of the high-redshift evolution of the intergalactic 
starlight photon field, including its spectral energy distribution out to
frequencies beyond the Lyman limit. To estimate this evolution, we have
followed a recent analysis of Fall, Charlot \& Pei, 
which reproduces the redshift dependence of the starlight 
background emissivity obtained by the Canada-France
redshift survey group. 
We also include the UV background from quasars.

We give our results for the \gray opacity as a function
of redshift out to a redshift of 3. We also give predicted \gray spectra 
for selected blazars and extend our calculations of the 
extragalactic \gray background from blazars to an energy of 500 GeV with
absorption effects included. Our results indicate that the extragalactic
\gray background spectrum from blazars should steepen significantly 
above 20 GeV, owing to extragalactic absorption. Future observations of
a such a steepening would thus provide a test of the blazar origin
hypothesis for the \gray background radiation. We also note that our
absorption calculations can be used to place limits on the redshifts
of \gray bursts; for example, our calculated opacities indicate that the 
17 Feb. 1994 burst observed by {\it EGRET} must
have originated at $z \le \sim 2$. Finally, our estimates of the high-energy 
\gray background
spectrum are used to determine the observability of multi-GeV \gray
lines from the annihilation of supersymmetric
dark-matter particles in the galactic halo.

\end{abstract}

\section{Introduction}

The {\it EGRET} experiment aboard the {\it Compton Gamma Ray Observatory}
has detected more than 50 blazars extending out to redshifts greater than
2 (Thompson, \etal\ 1996). It is expected that \grays from 
blazars with energies above the
threshold energy for electron-positron pair production through interactions
with low energy intergalactic photons will be annihilated, cutting off
the high energy end of blazar spectra. Such absorption is strongly dependent
on the redshift of the source (Stecker, De Jager \& Salamon 1992). 
Stecker \& De Jager (1997) have calculated the absorption of extragalactic
\grays above 0.3 TeV at redshifts up to 0.54 and presented a comparison
with the spectral data for the low redshift blazar Mrk 421. 

The study of blazar spectra at energies below 0.3 TeV is a more complex and
physically interesting subject. In addition to intergalactic absorption, one
must be able to distinguish and to separate out the effects of intrinsic 
absorption and natural cutoff energies in blazar emission spectra. 
Initial estimates of intergalactic absorption of 10 to 300 GeV \grays 
in blazar spectra at higher redshifts have been given by Stecker (1996), 
Stecker \& de Jager (1996) and Madau \& Phinney (1996). However, in order
to calculate such high-redshift absorption properly, it is
necessary to determine the spectral distribution of the intergalactic low 
energy photon background radiation as a function of redshift as realistically 
as possible. This calculation,
in turn, requires observationally based information on the evolution of the 
spectral energy distributions (SEDs) of IR through UV starlight from galaxies,
particularly at high redshifts. Conversely, observations of 
high-energy cutoffs in the
\gray spectra of blazars as a function of redshift, which may enable one to 
separate out intergalactic absorption from redshift-independent cutoff 
effects, could add to our knowledge of galaxy formation and early galaxy 
evolution. 
In this regard, it should be noted that the study of blazar spectra in the 
10 to 300 GeV range is one of the primary goals of a next generation
space-based \gray telescope, {\it GLAST (Gamma-ray Large Area Space Telescope)}
(Bloom 1996) as well as a number of ground-based \gray telescopes 
currently under construction.
	
\section{Redshift Dependence of the Intergalactic Low Energy SED}

Our main goal is to calculate the opacity of intergalactic space to high energy
\grays as a function of redshift.  This depends upon the 
number density of soft target photons 
(IR to UV) as a function of redshift,
whose production is dominated by stellar emission.  To evaluate the
SED of the IR-UV intergalactic radiation field
we must integrate the total stellar emissivity 
over time.  This requires an
estimate of the dependence of stellar emissivity on redshift. Previous work
has either assumed that all of the background was in place at high redshifts, corresponding to a burst of star formation at the initial redshift 
(Stecker 1996; Stecker \& De Jager 1996; MacMinn and Primack 1996) or 
strong evolution (similar to a burst), or no evolution (Madau and Phinney 
1996). In this paper, we use a more realistic model which is consistent with 
recent observational data.

\subsection{Basic Calculation of Stellar Emissivity} 

Pei \& Fall (1995) have devised a method for calculating stellar emissivity
which bypasses the uncertainties associated with estimates of poorly defined
luminosity distributions of evolving galaxies.
The core idea of their 
approach is to relate the star formation rate 
directly to the evolution of the neutral gas density in damped
\lya systems, and then to use stellar population synthesis models to
estimate the mean co-moving stellar emissivity ${\cal E}_{\nu}(z)$
(erg/s-cm$^{3}$-Hz)
of the universe as a function of frequency $\nu$ and
redshift $z$ (Fall, Charlot \& Pei 1996).
Our calculation of stellar emissivity closely follows this 
elegant analysis, with minor modifications as described below.

Damped \lya systems are high-redshift clouds of gas whose neutral
hydrogen surface density is large enough ($>2\times 10^{20}$ cm$^{-2}$)
to generate saturated \lya
absorption lines 
in the spectra of background quasars that 
happen to lie along and
behind  common lines of sight to these clouds.  
These gas systems are believed
to be either precursors to galaxies or young galaxies themselves, 
since their neutral hydrogen (HI)
surface densities are comparable to those of spiral galaxies today, and their
co-moving number densities are consistent with those of present-day galaxies
(Wolfe 1986; see also Peebles 1993).  It is in these systems that initial
star formation presumably took place, so there is a relationship between
the mass content of stars and of gas in these clouds; if there is no infall
or outflow of gas in these systems, the systems are ``closed'', so that the
formation of stars must be accompanied by a reduction in the neutral gas
content.  Such a variation in the HI surface densities of
\lya systems with redshift is seen, and is used by Pei \& Fall (1995)
to estimate the mean cosmological rate of star formation back to 
redshifts as large as $z=5$.

Pei \& Fall (1995) have estimated the
neutral (HI plus HeI) co-moving gas density 
$\rho_{c}\Omega_{g}(z)$ in damped \lya systems 
from observations of the redshift evolution of 
these systems by Lanzetta, Wolfe, \& Turnshek (1995).  (Here
$\rho_{c}=3H_{0}^{2}/8\pi G$
is the critical mass density of the universe.
The deceleration parameter is assumed throughout 
to be $q_{0}=0.5$, with cosmological constant $\Lambda=0$.)
Lanzetta, \etal\ have observed 
that while the number density of damped \lya systems 
appears to be relatively constant over redshift, the fraction of higher
density absorption systems within this class of objects
decreases steadily with decreasing redshift.  They attribute this to a
reduction in gas density with time, roughly of the form
$\Omega_{g}(z)=\Omega_{g0}e^{z}$, where 
$\rho_{c}\Omega_{g0}$ is the 
current gas density in galaxies.  Pei \& Fall (1995) have
taken account of self-biasing effects to obtain a corrected
value of $\Omega_{g}(z)$; we have reproduced
their calculations to obtain $\Omega_{g}(z)$ under the assumptions that
the asymptotic, high redshift value of 
the neutral gas mass density is $\Omega_{g,i}=
1.6\times10^{-2}h_{0}^{-1}$, where $h_{0}\equiv H_{0}/(100$ km/s-Mpc).
In a ``closed galaxy'' model, the change in co-moving stellar mass density 
$\rho_{c}\dot{\Omega}_{s}(z)=-\rho_{c}\dot{\Omega}_{g}(z)$, since the gas 
mass density $\rho_{c}\Omega_{g}(z)$ is
being converted into stars. This determines the star formation
rate and consequent stellar emissivity (Pei \& Fall 1995).

To determine the mean stellar emissivity from the star
formation rate, an initial mass function (IMF) $\phi(M)$
must be assumed for the distribution of stellar masses $M$ in a
freshly synthesized stellar population. To further specify the luminosities
of these stars as a function of mass $M$ and age $T$, 
Fall, Charlot, \& Pei (1996) use the Bruzual-Charlot (BC)
population synthesis models
for the spectral evolution of stellar populations (Bruzual \& Charlot 1993,
Charlot \& Bruzual 1991).  
In these population synthesis models, the
specific luminosity
$L_{\rm star}(\nu,M,T)$ (erg/s-Hz), of a star of mass $M$ and age
$T$ is integrated over a specified IMF to obtain a total specific
luminosity $S_{\nu}(T)$ per unit mass (erg/s-Hz-g)
for an entire population,
in which all stellar members are produced 
simultaneously ($T=0$).  Following Fall, Charlot, and Pei (1996),
we have used in our calculations the BC model
corresponding to a Salpeter IMF,
$\phi(M)\,dM\propto M^{-2.35}\,dM$, where $0.1M_{\odot}<M<125M_{\odot}$.

The mean co-moving emissivity ${\cal E}_{\nu}(t)$
is then obtained by convolving over time $t$ 
the specific luminosity $S_{\nu}$ with 
the mean co-moving mass rate of star formation, $\rho_{c}\dot{\Omega}_{f}$:
\begin{equation}
\label{emissivity.eq}
{\cal E}_{\nu}(t)=\rho_{c}\int_{0}^{t}dt^{\prime}\,
S_{\nu}(T=t-t^{\prime})\dot{\Omega}_{f}(t^{\prime}).
\end{equation}
Note that the star mass formation rate $\rho_{c}\dot{\Omega}_{f}(t)$ 
that appears
in this equation is not the same as $\rho_{c}\dot{\Omega}_{s}(t)$,
the change in total stellar mass density.
This is because $\rho_{c}\dot{\Omega}_{s}$ is 
the rate at which mass is {\it permanently} being converted into stars;
since some stellar mass is continuously being
returned to the interstellar medium (ISM), the {\it instantaneous} mass
rate of star formation $\rho_{c}\dot{\Omega_{f}}$ is larger than 
$\rho_{c}\dot{\Omega}_{s}$, the two being related by
\begin{equation}
\label{massconversion.eq}
\dot{\Omega}_{s}(t)=\dot{\Omega}_{f}(t)
-\int_{0}^{t}dt^{\prime}\,\dot{R}(t-t^{\prime})
\dot{\Omega}_{f}(t^{\prime}),
\end{equation}
where $R(T)$, provided by the BC models, is the fraction 
of the initial mass of a generation of stars
formed at $T=0$ that has been returned to the ISM.

\subsection{Metallicity Corrections}

The BC models' specific luminosities $S_{\nu}(T)$ 
are calculated assuming that the metallicity 
content {\it Z} during star formation is fixed at our current solar
metallicity value ($Z_{\odot}=0.0169$).  However, the metallicity
content of the universe is not static, but evolves with redshift as
early populations of stars return freshly synthesized metals to the
interstellar medium during their various phases of mass loss.
For example, in a survey of 1/3 of the known damped
Lyman-alpha absorbers, Pettini {\it et al.} (1994) found that the typical
metallicity is 0.1 that of the present solar value at a redshift of
$z\approx 2$.
Since the specific luminosity of a star of a given mass is also
a function of its metallicity content (lower metallicities give
bluer spectra), the metallicity of a stellar population must be taken
into account when integrating the mean emissivity over redshift.

The effect of metallicity content in stellar population models has been
examined by Worthey (1994).
Using the IMF $\phi(M)\,dM\propto M^{-2.35}dM$ with
$0.1M_{\odot}<M<2M_{\odot}$, Worthey has calculated the mass-to-light ratios
$\left<M/L\right>$ as a function of population age $T$ and metallicity $Z$,
for the color bands $U$ through $M$.
We have plotted his $\left<M/L\right>$ values for the $U$ and $B$
bands in Figures 1 and 2 respectively.
One can see that for a fixed metallicity, 
the logarithm of the luminosity decreases
approximately linearly with the logarithm of population age, and that 
for a fixed age the $U$ and $B$ luminosities decrease
as the metallicity increases.
We have made a linear fit to each fixed-metallicity $\left<M/L\right>(T)$ 
computed data set,
obtaining a metallicity correction factor 
factor ${\cal L}_{X}(Z/Z_{\odot})=
L_{X}(Z)/L_{X}(Z_{\odot})$, where $X$ designates the color band.
>From the parallel linear fits made to the computed data
for each $y\equiv\log (Z/Z_{\circ})$ value, 
it is seen that a common correction factor, ${\cal L}$,
applied to each $y\ne 0$ data set will bring these data into rough
agreement with the $y=0$ values of $\left<M/L\right>(T)$.  These correction factors
are plotted in the inset figures, whose abscissa is $Z/Z_{\odot}$ and
whose ordinate is the correction factor.  Our fit to Worthey's computed
data in Figure 1 gives a continuous correction factor ${\cal L}(Z)$ for
$\lambda=0.35$ $\mu$m, the center of the $U$ band.  Similar fits to the
$B$ band data (Figure 2) and to the $V$, $I$, and $K$ band data
(not shown) result in the relation

\begin{equation}
\label{worthey.eq}
\log{\cal L}(\lambda,Z)\equiv\log\left[\frac{L(\lambda,Z)}
{L(\lambda,Z_{\odot})}\right]\approx\left[0.33-\frac{0.30}{\lambda}
\right]\log\left(\frac{Z}{Z_{\odot}}\right) +
\left[0.066-\frac{0.063}{\lambda}\right]\left[
\log\left(\frac{Z}{Z_{\odot}}\right)\right]^{2},
\end{equation}
for $0.3\ \mu$m $<\lambda<2.2\ \mu$m.
Outside of this wavelength region we take 
${\cal L}(\lambda<0.3\mu{\rm m},Z)={\cal L}(\lambda=0.3\mu{\rm m},Z)$
and ${\cal L}(\lambda>2.2\mu{\rm m},Z)={\cal L}(\lambda=2.2\mu{\rm m},Z)$.
Note that increased metallicity gives a redder population spectrum
(Bertelli \etal\ 1994).

%\begin{figure}[h]
%\label{uband.fig}
%\epsfysize=3in
%\epsfbox{worthyufinal.ps}
%\caption{

%}
%\end{figure}

%\begin{figure}[h]
%\label{bband.fig}
%\epsfysize=3in
%\epsfbox{worthybfinal.ps}
%\caption{

%}
%\end{figure}

Limitations to this correction factor include the fact that
Worthey's calculations only apply to stars with ages greater than $T=1.5$ Gyr,
and that the
upper mass limit of his IMF ($2M_{\odot}$) is much lower than that
of the BC model which we employ ($125M_{\odot}$).
Additional uncertainty exists below 0.3 $\mu$m since Worthey's
calculations extend only to the $U$ band.  We have chosen to 
assume a constant enhancement factor below $\lambda=0.3 \mu$m.
For all of the above reasons, our enhancement
factor ${\cal L}$ is really a conservative lower limit to the corrections
to the BC models in the ultraviolet. 

Population synthesis models in which varying metallicity is included
do exist (Bertelli \etal\ 1994), and efforts to reconcile differences
in computed spectra generated by these various models have been made
(Charlot, Worthey, and Bressan, 1996).

\subsection{Absorption by Dust and Gas within Source}

The emissivity ${\cal E}_{\nu}$ given in Eq. \ref{emissivity.eq} assumes that
all stellar emission escapes from the gas system which contains the
stars.  However, some absorption of stellar
radiation occurs both by dust and gas within the larger damped \lya
systems.
Above the Lyman limit, this absorption is dominated by dust, while
below the Lyman limit, absorption by neutral hydrogen and singly-ionized
helium dominates.  Defining the mean
transmission fractions, averaged over the optical depths of damped \lya
systems, by 
${\cal T}_{\rm dust}(\nu,z)$ and 
${\cal T}_{\rm gas}(\nu,z)$, the final expression for
the effective stellar emissivity is
\begin{equation}
\label{atten_emiss.eq}
{\cal E}_{\nu}(z)={\cal T}_{\rm dust}(\nu,z)
{\cal T}_{\rm gas}(\nu,z)\rho_{c}
\int_{z}^{z_{\rm max}}\frac{dz^{\prime}}{H_{0}(1+z^{\prime})^{2.5}}
\dot{\Omega}_{f}(z^{\prime}){\cal L}(\nu,z^{\prime})
S_{\nu}[T=t(z)-t(z^{\prime})].
\end{equation}

The distribution of optical depths $\tau_{d}$ of \lya clouds due to
dust is
can be adequately represented by
$f(x)=f_{0}x^{-\alpha}e^{-x}$, where $x=\tau_{d}/\tau_{\ast}(z)$,
$\tau_{*}(z)$ being a characteristic (redshift dependent)
cloud dust opacity, and $\alpha\approx 1$ (Fall \etal\ 1996).  Under the
assumption that both dust and stars are uniformly distributed throughout each
\lya cloud, the fraction of radiation 
${\cal T}^{(1)}(\tau_{d})$ produced by stars in a given cloud of optical depth
$\tau_{d}$
that escapes dust absorption is given by
\begin{equation}
\label{twostream.eq}
{\cal T}^{(1)}(\tau)=\frac{2}{\tau\eta}\left[\frac{1-e^{-\tau\eta}}
{(1+\eta)+(1-\eta)e^{-\tau\eta}}\right],
\end{equation}
where $\eta=\sqrt{1-\omega}$, and $\omega$ is the average albedo of dust,
taken to be the same value as in our Galaxy ($\omega\sim 0.4$ to 0.6; 
Whittet 1992).  (We have calculated Eq. \ref{twostream.eq} using the
2-stream approximation [Chandrasekhar 1950]). 
We note that the dust opacity
$\tau_{d}$ in Eq.\ref{twostream.eq} 
is assumed to be proportional to the HI surface column density 
$N_{ \rm HI}$ and metallicity $Z$,
\begin{equation} \label{C}
\tau_{d}(\nu,z)=\frac{Z(z)}{Z_{\odot}}
\left(\frac{N_{ \rm HI}}{6.3\times10^{21}{\rm cm}^{-2}}
\right)E(\nu),
\end{equation}
where $E(\nu)$ is the normalized galactic interstellar dust
extinction
curve (Savage and Mathis 1979).
Integrating Eq.\ref{twostream.eq}
over the \lya opacity distribution function $f$ of Pei \& Fall (1995),
we obtain ${\cal T}_{\rm dust}(\nu,z)$, and find it to have a minor effect
on the emissivity, ${\cal T}_{\rm dust}(\nu,z)$ being typically of order unity.

Below the Lyman limit ($\lambda < 0.0912\ \mu$m), the opacity is dominated
by neutral gas absorption: $\tau_{g}(\nu)=N_{\rm HI}\sigma_{\rm HI}(\nu) +
N_{\rm HeI}\sigma_{\rm HeI}(\nu)$, where $\sigma_{\rm HI}$ and 
$\sigma_{\rm HeI}$ are the HI and HeI photoionization cross sections
(Osterbrock 1989).  With the $N_{\rm HI}$ and $N_{\rm HeI}$
distributions of the \lya systems being related to the dust opacity 
distribution $f(\tau_{d})$ through Eq.\ref{C}, the distribution for
$\tau_{g}$ can be obtained.  Integrating Eq.\ref{twostream.eq} (now with
$\eta=1$), weighted with the $\tau_{g}$ distribution, gives
${\cal T}_{\rm gas}(\nu,z)$.

\subsection{Numerical Results}

Figure 3
shows the calculated stellar
emissivity as a function of redshift at 0.28 $\mu$m, 0.44 $\mu$m,
and 1.00 $\mu$m, both with and without the metallicity correction
factor ${\cal L}(\lambda, Z)$.
We have also plotted the observations of the cosmic
emissivity by the Canada-French
Redshift Survey (Lilly, Le Fevre, Hammer, \& Crampton 1996) at these
rest-frame wavelengths for comparison.  With a lower mass cutoff of
$0.1 M_{\odot}$ in the IMF, we obtain emissivities which are roughly
a factor of 2 higher than those obtained by Lilly, {\it et al.} (1996).  
To bring our
emissivities down to the observed values requires that we reduce the
lower mass limit in the IMF to $0.02 M_{\odot}$, which puts a fraction
(0.45) of the mass into effectively nonluminous compact objects.  We
note that a similar reduction was achieved by Fall, \etal\ (1996)
by modifying the power
law index in the IMF; a higher index results in a lower emissivity
(Pei 1996, personal communication).  

Overall, our emissivities, both with and without the metallicity
corrections, are in reasonable agreement with 
the data at lower redshifts (Lilly, \etal\ 1996). 
Although the differences for ${\cal E}_{\nu}$ between the no-metallcity 
and metallicity cases for $z<1$ are not great, {\it they become substantial at
larger redshifts for both optical and UV wavelengths}.  This has notable effects
on the opacity of the radiation background to high energy \grays, as
will be seen in Section 4.  We note that
our dotted-line curves in Figure 3 (no metallicity correction) are essentially
a reproduction of the emissivities calculated by Fall, \etal\ (1996).

In all cases as shown in Figure 3, the stellar emissivity in the universe peaks
at $ 1 \le z \le 2$, dropping off at both lower and higher redshifts. Indeed,
Madau, \etal\ (1996) have used observational data from the Hubble Deep Field to
show that metal production has a similar redshift distribution, such production
being a direct measure of the star formation rate. (See also the review by 
Madau (1996).)

%\begin{figure}[h]
%\label{emissvsz.fig}
%\epsfysize=3in
%\epsfbox{emissvszfinal.ps}
%\caption{

%}
%\end{figure}

\subsection{Calculation of the Diffuse Radiation Energy Density}
\label{energydensity}

The co-moving radiation energy density $u_{\nu}(z)$ 
(erg/cm$^{3}$-Hz) is the
time integral of the co-moving emissivity ${\cal E}_{\nu}(z)$,
\begin{equation} \label{B}
u_{\nu}(z)=
\int_{z}^{z_{\rm max}}dz^{\prime}\,{\cal E}_{\nu^{\prime}}(z^{\prime})
\frac{dt}{dz}(z^{\prime})e^{-\tau_{\rm eff}(\nu,z,z^{\prime})},
\end{equation}
where $\nu^{\prime}=\nu(1+z^{\prime})/(1+z)$ and $z_{\rm max}$ is the
redshift corresponding to initial galaxy formation.
The extinction term $e^{-\tau_{\rm eff}}$ 
accounts for the absorption of ionizing photons by the clumpy
intergalactic medium (IGM) that lies between the source and observer;
although the IGM is effectively transparent to non-ionizing photons,
the absorption of photons by HI,
HeI and HeII can be considerable (Madau 1995).
The presence of damped \lya and Lyman-limit systems (Lanzetta, \etal\ 
1995) and the Lyman-alpha forest, coupled
with the absence of a HI Gunn-Peterson effect (Gunn \& Peterson 1965;
Steidel \& Sargent 1987) indicates that essentially all of the HI, HeI,
and HeII exists within intergalactic clouds whose measured HI column 
densities range from approximately $10^{13}$ to $10^{22}$ cm$^{-2}$.

The effective optical depth $\tau_{\rm eff}$ between a source at
redshift $z^{\prime}$ and an observer at redshift $z$ owing to 
Poisson-distributed
intervening Lyman-alpha clouds is given by (Paresce, McKee, \&
Bowyer 1980) 
\begin{equation} \label{E}
\tau_{\rm eff}(\nu,z,z^{\prime})=
\int_{z}^{z^{\prime}}dz^{\prime\prime}\,\int_{0}^{\infty}
dN_{\rm HI}\,\frac{\partial^{2}N}{\partial N_{\rm HI}
\partial z^{\prime\prime}}(1-e^{\tau(\nu^{\prime\prime})}),
\end{equation}
where $\tau(\nu)=\left[ N_{\rm HI}\sigma_{\rm HI}(\nu) +
N_{\rm HeI}\sigma_{\rm HeI}(\nu) + 
N_{\rm HeII}\sigma_{\rm HeII}(\nu)\right]$, 
$\nu^{\prime\prime}=\nu(1+z)/(1+
z^{\prime\prime})$, and $\frac{\partial^{2}N}{\partial N_{\rm HI}
\partial z}$ is the distribution function of clouds in redshift $z$
and column density $N_{\rm HI}$.  As pointed out by Madau \& Shull (1996),
when $\tau\ll 1$, $\tau_{\rm eff}$ is just the mean optical depth of the
clouds; when $\tau\gg 1$, $\tau_{\rm eff}$ becomes the number of
optically thick clouds between the source and observer, so that the
Poisson probability of encountering no thick clouds is $e^{-\tau_{\rm eff}}$,
as required.

For the distribution function of Lyman-alpha clouds we use the
parameterization of Madau (1995) (see also Miralda-Escud\'{e} \&
Ostriker, 1990, Model A2):
\begin{equation} \label{F}
\frac{\partial^{2}N}{\partial N_{\rm HI}\partial z}=\left\{
\begin{array}{ll}
2.4\times10^{7}N_{\rm HI}^{-1.5}(1+z)^{2.46}, & 
	2\times10^{12} < N_{\rm HI} < 1.59\times10^{17} {\rm cm}^{2} \\
1.9\times10^{8}N_{\rm HI}^{-1.5}(1+z)^{0.68} &
	1.59\times10^{17} < N_{\rm HI} < 10^{20} {\rm cm}^{-2}.
\end{array} \right.
\end{equation}
Using Eqs. \ref{E} and \ref{F} and the 
stellar emissivity ${\cal E}_{\nu}(z)$ in Eq.\ref{B}, we obtain
the background energy density $u_{\nu}(z)$, shown in Figures 4 and 5,
%\ref{uvdens_with.fig} and \ref{uvdens_without.fig}, 
calculated with
and without the metallicity correction, ${\cal L}$, respectively.
These also include the contribution to the UV
background from QSOs (Madau 1992), which are believed to dominate the diffuse
background radiation below the Lyman limit and to be responsible
for the early ($z>5$; see Schneider, Schmidt, \& Gunn 1991) 
reionization of the IGM.  

Although it is possible that
UV emission from QSOs alone may be able to account for the nearly complete
reionization of the IGM (Meiksin \& Madau 1993; Fall \& Pei 1993;
Madau \& Meiksin 1994), it has been argued that additional sources of
of ionizing radiation are required (Miralda-Escud\'{e} \& Ostriker 1990),
these perhaps being young galaxies which leak a fraction (up to $\sim 15\%$)
of their ionizing radiation through HII ``chimneys'' (Dove \& Shull 1994;
Madau \& Shull 1996).  We have therefore assumed in our calculations that 15\%
of the stellar emission escapes from the galaxies (protogalaxies) through 
these chimneys, unattentuated by dust or gas.
(We note, however, that recent observations of four starburst galaxies
by the Hopkins UV Telescope (Leitherer {\it et al.} 1995) indicate
that less than 3\% of Lyman continuum photons escape from these sources.)
Figures 4 and 5
%\ref{uvdens_with.fig} and \ref{uvdens_without.fig} 
indicate that
in our calculation the $\lambda < 0.0912 \mu$m background is indeed
dominated by QSOs, so that the actual value of the escape fraction 
we choose is not too significant.

The intergalactic energy densities given in Figures 4 and 5 are quite 
consistent with the present upper limits in the UV (Martin \& Bowyer
1989; Mattila 1990; Bowyer 1991; Vogel, Weymann, Rauch \& Hamilton 1995).
It should be noted that our results as shown in figures 4 and 5 give 
emissivities {\it from starlight only} and do not include dust emissivities in
the mid-infrared and far-infrared.  

%\begin{figure}[h]
%\label{uvdens_with.fig}
%\epsfysize=3in
%\epsfbox{uvdens_with.ps}
%\caption{

%}
%\end{figure}

%\begin{figure}[h]
%\label{uvdens_without.fig}
%\epsfysize=3in
%\epsfbox{uvdens_without.ps}
%\caption{
%}
%\end{figure}

\section{Opacity of the Radiation Background and its Effect on
Blazar Spectra}
%\label{opacity}

With the co-moving energy density $u_{\nu}(z)$ evaluated, the optical
depth for \grays owing to electron-positron pair production 
interactions with photons of the stellar radiation
background can be determined from the expression
(Stecker, \etal\ 1992)
\begin{equation} \label{G}
\tau(E_{0},z_{e})=c\int_{0}^{z_{e}}dz\,\frac{dt}{dz}\int_{0}^{2}
dx\,\frac{x}{2}\int_{0}^{\infty}d\nu\,(1+z)^{3}\left[\frac{u_{\nu}(z)}
{h\nu}\right]\sigma_{\gamma\gamma}[s=2E_{0}h\nu x(1+z)^2],
\end{equation}
where $E_{0}$ is the observed \gray energy, $z_{e}$ is the redshift of
the \gray source, $x=(1-\cos\theta)$, $\theta$ being the angle between
the \gray and the soft background photon, $h$ is Planck's constant, and
the pair production cross section $\sigma_{\gamma\gamma}$ is zero for
center-of-mass energy $\sqrt{s} < 2m_{e}c^{2}$, $m_{e}$ being the electron
mass.  Above this threshold, 
\begin{equation} \label{H}
\sigma_{\gamma\gamma}(s)=\frac{3}{16}\sigma_{\rm T}(1-\beta^{2})
\left[ 2\beta(\beta^{2}-2)+(3-\beta^{4})\ln\left(\frac{1+\beta}{1-\beta}
\right)\right],
\end{equation}
where $\beta=(1-4m_{e}^{2}c^{4}/s)^{1/2}$.

Figures 6 and 7
%\ref{opac_with.fig} and \ref{opac_without.fig} 
show the opacity $\tau(E_{0},z)$ for the energy
range 10 to 500 GeV, calculated with and without the metallicity correction.
Extinction of \grays is negligible below 10 GeV.
Above 500 GeV, interactions with photons with wavelengths of tens of
$\mu$m become important, so that 
one must include interactions from infrared photons generated by dust
reradiation
(Stecker \& De Jager 1997), which
we have neglected here. For 300 GeV \grays, 
at redshifts below 0.5, our opacities agree with the
with the opacities obtained by Stecker
\& De Jager (1997). Our calculated opacity, even with the metallicity 
correction, is probably somewhat low in the 10 to 30 GeV energy range,
because we have underestimated the value of ${\cal L}$ in the UV
(see previous discussion).

Note that these calculated opacities are {\it independent} of the value
chosen for $h_{0}$, as seen in Eqs. 4, 7, and 10.  The emissivity
${\cal E}_{\nu}$ in Eq. 4 scales as $h_{0}^{2}$, since neither 
$S_{\nu}$, ${\cal L}$ nor $dt\, \dot{\Omega_{f}}$ depends on $h_{0}$,
while $\rho_{c}$ scales as $h_{0}^{2}$.  Eq. 7 shows then that 
$u_{\nu}$ scales as $h_{0}$, and in Eq. 10 this $h_{0}$ factor is
cancelled by the integration over time $t$.

%\begin{figure}[h]
%\label{opac_with.fig}
%\epsfysize=3in
%\epsfbox{opac_with.ps}
%\caption{

%}
%\end{figure}

%\begin{figure}[h]
%\label{opac_without.fig}
%\epsfysize=3in
%\epsfbox{opac_without.ps}
%\caption{

%}
%\end{figure}

With the \gray opacity $\tau(E_{0},z)$ calculated out to
$z=3$,
the cutoffs in blazar \gray spectra caused by extragalactic pair 
production interactions with stellar photons can be predicted.
Figure 8
%\ref{blazaratt.fig} 
shows the effect of the intergalactic radiation
background on a few of the \gray blazars (``grazars'') observed by {\it EGRET},
{\it viz.}, 1633+382, 3C279, 3C273, and Mrk 421.
We have assumed that the mean spectral indices obtained for these sources by
{\it EGRET} extrapolate out to higher energies 
attenuated only by intergalactic 
absorption.  Observed cutoffs in grazar spectra 
may be intrinsic cutoffs in \gray production in
the source, or may be caused
by intrinsic \gray absorption within the source itself.  Whether 
cutoffs in grazar spectra are 
primarily caused by intergalactic absorption can be determined by
observing whether the grazar cutoff energies have the 
type of redshift dependence predicted here.

%begin{figure}[h]
%\label{blazaratt.fig}
%\epsfysize=3in
%\epsfbox{blazaratt.ps}
%\caption{

%}
%\end{figure}

Figure 8
%\ref{blazaratt.fig} 
indicates that the next 
generation of satellite and ground-based \gray detectors,
both of which will be designed to explore the energy range between 10
and 300 GeV, will be able to reveal information about low-energy radiation 
produced by galaxies at various redshifts and at different stages in their
evolution.

\section{Constraints on Gamma-ray Bursts}

Our opacity calculations have implications for the 
determination of the origin of \gray bursts, if such bursts are cosmological.
As indicated in Figure 6,
%\ref{opac_with.fig} and \ref{opac_without.fig}, 
\grays above an energy of $\sim$ 15 GeV will
be attenuated if they at emitted at a redshift of $\sim$ 3.
On 17 February 1994, the {\it EGRET} telescope observed a \gray burst
which contained a photon of energy $\sim$ 20 GeV (Hurley, \etal 1994).
If one adopts the opacity results which include our conservative 
metallicity correction
(Figure 6), this burst would be constrained to have originated at a redshift
less than $\sim$2. (An estimated redshift constraint of $\sim$ 1.5 was given 
by Stecker and De Jager (1996), based on a simpler model.)
Future detectors may be able to place redshift constraints on bursts 
observed at higher energies.

\section{The High Energy Gamma Ray Background from Blazars}

In a previous paper
(Stecker \& Salamon 1996), we presented a model for calculating the 
extragalactic \gray
background (EGRB) due to unresolved grazars.
We gave results for \gray energies up to 10 GeV (where there is 
effectively no \gray absorption) which were compared to
preliminary {\it EGRET} data (Kniffen \etal\ 1996)
Using the intergalactic \gray opacities calculated here, we can now extend
the results of this EGRB model out to an energy of 0.5 TeV.

Our EGRB model
assumes that the grazar luminosity function is related to
that of flat spectrum radio 
quasars (FSRQ), so that we can 
use FSRQ luminosity and redshift distributions
(Dunlop and Peacock, 1990) to obtain a grazar luminosity function.
The effects of grazar flaring states, \gray spectral index variation, and
redshift dependence are also been included in this model; see Stecker and
Salamon (1996) for details.
By integrating the grazar
luminosity function weighted by our new opacity results, we obtain a
grazar background spectrum up to 500 GeV which properly includes the effect of
\gray absorption.

Figure 9
%\ref{attgrb.fig}, 
shows this EGRB spectrum 
compared with the preliminary {\it EGRET} data. Note that the 
spectrum is concave at energies below 10 GeV, reflecting the 
dominance of hard-spectrum grazars at high energies and softer-spectrum
grazars at low energies; it then steepens 
above 20 GeV, owing to extragalactic absorption by pair-production 
interactions with radiation from external galaxies, particularly at
high redshifts. Both the concavity and the steepening are signatures
of a blazar dominated \gray background spectrum.

%\begin{figure}[h]
%\label{attgrb.fig}
%\epsfysize=3in
%\epsfbox{attgrb.ps}
%\caption{

%}
%\end{figure}

Because the extragalactic \gray background in our model is made up of a 
superposition of {\it lower-luminosity, unresolved} grazars, its intensity
is determined by the number of sources in the Universe which are below the
detection threshhold of a particular telescope. A telescope with a superior 
point source sensitivity gives a higher source count, 
thereby reducing the number of unresolved
sources which constitute the diffuse \gray background.  
In Figure 9, 
%\ref{attgrb.fig}, 
the
upper spectra which are close to the {\it EGRET} data are obtained using the 
{\it EGRET}
threshold; the lower curves correspond to the projected sensitivity of the
proposed next generation {\it GLAST} satellite detector, which is expected to 
have a detection threshhold of $\sim 2\times 10^{-9}$ cm$^{-2}$s$^{-1}$ above 
0.1 GeV. 

It should also be noted that above 10 GeV, blazars may have natural cutoffs in
their source spectra (Stecker, De Jager \& Salamon 1996) and intrinsic 
absorption may also be important in some sources (Protheroe 
\& Biermann 1996). Thus, above 10 GeV our 
calculated background flux from unresolved blazars, shown in Figure 9, may 
actually be an upper limit.

\section{Observability Above Background of a Multi-GeV Gamma-ray Line Produced by Neutralinos}

The nature of the dark matter in the universe is one of the most important
fundamental problems in astrophysics and cosmology.
The non-baryonic mixed dark matter model with a 
total $\Omega = 1$ (Shafi \& Stecker 1984) 
gave predictions for fluctuations in the cosmic background 
radiation (Schaefer, Shafi \& Stecker 1989; Holtzman 1989) which were 
found to be in good agreement with the 
later COBE measurements. The best agreement appears to be found for 
$\sim$ 20\% hot dark matter, of which massive neutrinos 
are the most likely candidates, and $\sim$ 80\% cold dark matter 
(Pogosian \& Starobinsky 1993, 1995; Ma \& Bertchinger 1994; Klypin, \etal\
1995; Primack, \etal\ 1995; Liddle, \etal\ 1996; Babu, Schaefer \& Shafi 1996).

The most popular cold dark matter
particle candidates are the lightest sypersymmetric particles
(LSPs), the neutralinos (hereafter designated as $\chi$ particles).
Cosmologically important $\chi$ particles must annihilate with a
weak cross section, $\langle\sigma v\rangle_{A} \sim$ 10$^{-26}$ cm$^3$s$^{-1}$; calculations show that such cross sections lead to a value for 
$\Omega_{\chi} \sim 1$ with $\Omega_{\chi}$$ \propto$$ \langle\sigma v\rangle_{A}^{-1}$. The fact that supersymmetry neutralinos are predicted to have
such weak annihilation cross sections is an important reason why they
have become such popular dark matter candidates. Preliminary LEP 2 results 
give a lower limit on the mass of the $\chi$ of
$M_{\chi} \geq 20$ GeV (Ellis, Falk, Olive \& Schmitt 1996).

In the minimal supersymmetry model (MSSM), $\chi$ can be generally described
as a superposition of two gaugino states and two Higgsino states. Grand 
unified models with a universal gaugino mass generally favor states where
$\chi$ is almost a pure B-ino ($\tilde{B}$) ({\it e.g.} Diehl, \etal\ 1995), 
but other states such as photinos and Higgsinos are generally allowed by the 
theory. Kane \& Wells (1996) have presented possible accelerator 
evidence from CDF that $\chi$ may be a Higgsino of mass $\sim$ 40 GeV. 
  
Dark matter neutralinos will produce \grays by 
mutual pair annihilation. This process is expected to occur because 
neutralinos are Majorana fermions, {\it i.e.}, they are their own 
antiparticles. Indeed, most of this mutual annihilation would have
occured in the very early universe, a process which determines the
present (``freeze out'') value of $\Omega_{\chi}$ and leads to the relation
$\Omega_{\chi} \propto \langle\sigma_{A} v\rangle^{-1}$, where the bracketed
quantity is the thermal-averaged annihilation cross section times velocity
(Ellis, \etal\ 1984). This leads to the relation that the \gray flux from
neutralino annihilation is inversely proportional to $\Omega_{\chi}$. 
Thus, the annihilation \gray flux is limited from below by cosmological 
constraints on the maximum value of $\Omega_{\chi}$. 
 
There are two types of \gray spectra produced by $\chi\chi$ annihilations,
{\it viz.}, (1) \gray continuum spectra from the decay of secondary
particles produced in the annihilation process, and (2) \gray lines, produced
primarily from the process $\chi\chi \to \gamma\gamma$ 
({\it e.g.},  Rudaz 1989).
The cosmic \gray flux from $\chi\chi$ annihilation is proportional to the
line-of-sight integral of the {\it square} of the $\chi$ particle density
times $\langle\sigma v \rangle_{A}$.

The continuum \gray production spectra from $\chi\chi$ annihilation
can be calculated for different types of neutralinos by starting with the 
appropriate branching ratios for annihilation into fermion-antifermion
pairs which produce hadronic cascades leading to the subsequent production 
and decay of neutral pions (Rudaz \& Stecker 1988; 
Stecker 1988; Stecker \& Tylka 1989). Stecker \& Tylka (1989) discuss in 
detail the various channels
involved in continuum \gray production via $\chi\chi$ annihilation and give 
the resulting spectra for some lower mass $\chi$ particles. Such continuum 
fluxes from $\chi\chi$ annihilations would be difficult to observe above 
the extrapolated cosmic background which we show in Figure 9.
However, with good enough sensitivity and energy resolution, it might be
possibile to observe a two-photon annihilation
line from $\chi\chi$ annihilation.
The general considerations for observability of this line were discussed
by Rudaz \& Stecker (1991). We update this discussion here, using (1) our
new calculation of the \gray background flux from blazars shown in Figure
9, (2) recent accelerator limits on supersymmetric particle masses, and (3)
the proposed sensitivity and energy resolution of a next generation
space based \gray telescope taken from the {\it GLAST} proposal (Elliot
1996).

The energy of the $\chi\chi \to \gamma\gamma$ decay line is $E_{\gamma} =
M_{\chi}$. The line width is given by Doppler broadening. For galactic
halo particles, this width is roughly $\beta_{\chi}M_{\chi}$ 
$\sim 10^{-3}M_{\chi}$, much smaller than the energy resolution proposed for
any future \gray telescope. 
Upper and lower limits on $\Omega$ yield lower and upper limits
on the \gray line flux respectively (see above). Other limits can be obtained 
in flux-energy space (Rudaz \& Stecker 1991). Accelerator determined 
lower limits on $M_{\chi}$ give lower limits on the line energy. 
Lower limits on the mass of the sfermion
exchanged in the annihilation process give upper limits on 
$\langle \sigma v \rangle_{A}$ 
since $\langle \sigma v \rangle_{A}$ $\propto$
$M_{\tilde{f}}^{-4}$. In fact, since the particle density $n_{\chi} =
\rho_{\chi}/M_{\chi}$ and $\langle \sigma v \rangle_{A}$ $\propto$
$M_{\chi}^2/M_{\tilde{f}}^4$, the predicted annnihilation line flux
$\phi(E_{\gamma}) \propto M_{\tilde{f}}^{-4}$.
Further limits are obtained from the inequality $M_{\tilde{f}} \ge M_{\chi}$,
which is the tautology following from the condition that $\chi$ be the LSP.

If we assume that annihilations occur mainly through slepton exchange, 
{\it i.e.}, $M_{\tilde{q}} \gg M_{\tilde{l}}$, we can obtain an upper limit 
on the 2$\gamma$ line flux. This is because LEP 1.5 gives a lower
limit of $\sim$ 70 GeV on the slepton mass (de Boer, Miquel, Pohl \& Watson 1996), whereas the substantially higher squark mass lower
limit of $\sim$ 150 GeV would imply much lower fluxes, since $\phi_{\gamma}$
$\propto M_{\tilde{f}}^{-4}$. 

The lower limit on the slepton mass implies an upper limit on the line flux 
from $\tilde{B}\tilde{B}$ annihilation such that the event rate for a next
generation \gray telescope with an aperture of 1 m$^2$sr would be
about 5 photons per year for a line in the energy range between 20 and 100 
GeV. (If the $\chi$ particles are Higgsinos, the event 
rate would be much lower.) Using all of these constraints, the allowed region
for a neutralino annihilation line in flux-energy space is plotted in Figure
10.

In constructing Figure 10, we have used the {\it GLAST} proposed estimate 
of the point source sensitivity after
a one-year full sky survey to estimate the background from unresolved faint 
blazars (see Figure 9 and the discussion in the previous section). 
We then obtain the background photon number for an appropriate exposure 
factor of 1 m$^2$yr-sr and energy resolution of 10\%, and plot the square 
root of this number, which represents 
the natural background fluctuations above which a line must be observed.
Of course, a higher exposure factor would reduce the point source background
and increase the sensitivity to a line flux, as would a better energy
resolution. It should also be noted that the background above 10 GeV shown in
Figures 9 and 10 may be overestimated (see previous section).

Another possible way in which dark matter may produce \grays and neutrinos is
if the LSP is allowed to decay to non-supersymmetric, ordinary particles.
Supersymmetry theories involve a multiplicative quantum number 
called {\it R-parity}, 
which is defined so that it is even for ordinary particles and odd for 
their supersymmetric
partners. Thus, if R-parity is conserved, as is usually assumed, 
the LSP is completely stable, making
it a potential dark matter candidate. However, such may not be the case.
R-parity may be very weakly broken, allowing the LSP to decay with branching
ratios involving \grays and neutrinos ({\it e.g.}, Berezinsky, Masiero \& 
Valley 1991). For $\chi$ particles to be the dark matter, their 
decay time should be considerably longer than the age of the universe.

The possible radiative decay $\chi \to \nu + \gamma$
will give a \gray line with energy $E_{\gamma} = M_{\chi}/2$. Such a
line has the potential of being more intense than the annihilation line.
Whereas the $\chi\chi$ annihilation rate and consequent line flux 
is cosmologically limited by requiring 
$\Omega_{\chi}$ to be a significant fraction near 1 (see previous discussion),
the decay-line flux is limited only by the particular physical supersymmetry 
model postulated and constraints from related accelerator and astrophysical 
data. Thus, invocation of $\chi$ decay involves a higher order of particle
theory model building and speculation. We only wish to mention here that
there is a possibility that a decay line may be sufficiently intense to be
observable above the background.

\section{Conclusions}

We have calculated the \gray\ opacity as a function of both energy and redshift
for redshifts as high as 3 by taking account of the evolution of both the
SED and emissivity of galaxies with redshift. In order to accomplish this, we 
have adopted the recent analysis of Fall, \etal\ (1996) and 
have also included the
effects of metallicity evolution on galactic SEDs. We have then considered the
effects of the \gray opacity of the universe on \gray bursts, blazar spectra,
and on the extragalactic \gray background from blazars. In particular, we
find that the 17 Feb. 1994 {\it EGRET} burst probably originated at $z \le 2$.
Because the stellar emissivity peaks between a redshift of 1 and 2, the 
\gray opacity which we derive shows little increase at higher redshifts. This
weak dependence indicates that the opacity is not determined by the initial
epoch of galaxy formation, contrary to the speculation of MacMinn and Primack
(1996).

The extragalactic \gray background, which can be accounted for as a 
superposition of spectra of unresolved blazars, and which we have predicted
to be concave between 0.03 and 10 GeV (Stecker \& Salamon 1996), should
steepen significantly above 20 GeV owing to our estimates of extragalactic
\gray absorption at moderate to high redshifts. Both the predicted concavity 
and steepening may be too subtle to detect with present data from {\it EGRET}.
However, next generation \gray telescopes which are presently being designed,
such as {\it GLAST}, may be able to observe these features and thereby test
the blazar background model. We also discuss the possible observability of
dark matter lines in the multi-GeV region, given our predicted extragalactic
background.

\acknowledgements {We thank Yichuan Pei and Matthew Malkan 
for very helpful conversations.}

\newpage

{\bf Figure Captions}

Figure 1: Plot of $U$-band mass-to-light ratios $M/L$ versus stellar
system age $T$ for various values of metallicity $Z$, taken from 
Worthey (1994), Table 5A.  The metallicity values are given on the
right-hand side of the figure as $y\equiv\log(Z/Z_{\odot})$, where $Z_{\odot}$
is solar metallicity.  Filled (open) squares 
correspond to the tabulated data for $y=+0.5$ ($y=+0.25$),
filled (open) circles for $y=0.0$ ($y=-0.22$), filled (open) triangles for
$y=-0.5$ ($y=-1.0$), filled inverted triangles for $y=-1.5$, and
open diamonds for $y=-2.0$.

Figure 2: Plot of the $B$-band mass-to-light ratios $M/L$ versus
stellar system age $T$ for various values of metallicity $Z$.  See
the caption of Figure 1 for details.

Figure 3: Emissivity as a function of redshift, calculated using Eq. 
\ref{atten_emiss.eq}, for three wavelength values, $\lambda=$0.28,
0.44, and 1.0 $\mu$m, for a Hubble constant value of $h_{0}=0.5$.  Note
that the emissivity scales as $h_{0}^{2}$ ({\it cf.} Eq. 4 and Section 3).
Solid line curves are for the case where the
metallicity correction factor (${\cal L}$ from eq. \ref{worthey.eq}) is used; 
dashed lines
give the emissivity when this correction factor is {\it not} included.  
The data from the Canada-French Redshift Survey (Lilly \etal\ 1996) 
are also plotted.

Figure 4: The intergalactic radiation energy density from stars and QSOs
as a function of wavelength for redshifts $z$ of 0, 1, 2, and 3, for a
Hubble constant value of $h_{0}=0.5$ (the energy density scales as
$h_{0}$; see Eq. 7 and Section 3).
These densities are calculated with the metallicity correction factor,
${\cal L}$, included.  

Figure 5: The same as Figure 4,
%\ref{uvdens_with.fig}, 
except {\it without} the metallicity correction factor.

Figure 6: The opacity $\tau$ of the universal soft photon background to
\grays as a function of \gray energy and source redshift.  These curves
are calculated with the metallicity correction factor 
included in the expression for stellar emissivity.  As discussed in the
text, these results are independent of the value chosen for $h_{0}$.

Figure 7: The same as Figure 6,
%\ref{opac_with.fig}, 
except {\it without}
the metallicity correction factor.

Figure 8: The effect of intergalactic absorption by pair-production on 
the power-law spectra of
four prominent grazars: 1633+382, 3C279, 3C273, and Mrk 421.  The
solid (dashed) curves are calculated with (without) the metallicity
correction factor.

Figure 9: The extragalactic \gray background energy spectrum from unresolved
grazars. The top and bottom sets of curves correspond to point-source 
sensitivities of $10^{-7}$ and $2\times 10^{-9}$ cm$^{-2}$s$^{-1}$,
respectively, for \gray energies above 0.1 GeV, corresponding
to the approximate point-source sensivities of the 
{\it EGRET} and {\it GLAST} detectors respectively.  Because the
FSRQ luminosity fuction that we employ 
scales as $h_{0}^{3}$ (Dunlop and Peacock, 1990), our calculated
EGRB spectrum scales as $h_{0}^{2}$ (see Eq. 10 in Stecker and Salamon,
1996).

Figure 10. The dot-dash polygon shows the allowed region of expected 
\gray photon counts calculated for a Bino ($\tilde{B}$) 
annihilation line as a function of $M_{\tilde{B}}$.
Present accelerator and cosmological constraints are indicated by the labels
on the sides of the polygon (see text). In the figure labels, the letter ``B''
represents the Bino ($\tilde{B}$) and the letter ``l'' represents the
slepton ($\tilde{\ell}$). As an illustration of how to read the figure, the 
arrow within the polygon indicates the line flux
upper limit for a Bino of mass 100 GeV. An exposure factor
of 1 m$^2$sr yr and an energy resolution of 10\% are assumed. We also show the
background fluctuation count rate appropriate to these parameters for the 
lower set of flux curves ({\it i.e.} with and without the metallicity 
correction) shown in Figure 9 (see text).


\begin{thebibliography}{}



\bibitem{Babu96} Babu, K.S., Schaefer, R.K. \& Shafi, Q. 1996, Phys. Rev., D53 
                 606
\bibitem{Bere91} Berezinsky, V., Masiero, A. \& Valle, J.W.F. 1991, Phys. 
                 Lett., B266, 382
\bibitem{Bert94} Bertelli, G., \etal\ 1994, A\&AS, 106, 275
\bibitem{Bl96} Bloom, E.D. 1996, Space Sci. Rev. 75, 109
\bibitem{Bo91} Bowyer, S. 1991, ARA\&A, 29, 59
\bibitem{br93} Bruzual A.,G. and Charlot, S. 1993, \apj, 405, 538
\bibitem{ch50} Chandrasekhar, S. 1950, {\it Radiative Transfer},
	(Clarendon Press, Oxford)
\bibitem{ch91} Charlot, S., and Bruzual A.,G. 1991, \apj, 457, 625
\bibitem{ch96} Charlot, S., Worthey, G., and Bressan, A. 1996,
	\apj, 457, 625.
\bibitem{de96} de Boer, W., Miquel, R., Pohl, M. \& Watson, N. 1996,
        {\it Joint Particle Physics Seminar, LEP Experiments Committee,
        CERN}.
\bibitem{Dieh95} Diehl, E. {\it et al.} 1995, Phys. Rev., D52, 4223
\bibitem{do94} Dove, J.B. and Shull, J.M. 1994, \apj, 430, 222
\bibitem{du90} Dunlop, J.S. and Peacock, J.A. 1990, MNRAS, 247, 19
\bibitem{Elli84} Ellis, J., {\it et al.} 1984, Nucl. Phys., B238, 453
\bibitem{el96} Ellis, J., Falk, T., Olive, K.A. \& Schmitt, M. 1966, 
        CERN preprint.
\bibitem{fa96} Fall, S.M., Charlot, S., and Pei, Y.C. 1996, \apj, 464, L43
\bibitem{fa93} Fall, S.M. and Pei, Y.C. 1993, \apj, 402, 479
%\bibitem{fi95} Fichtel, C.E., 1995, {\it Proc 3rd Compton Observatory 
%Symposium}, A \& A S, in press 
\bibitem{gu65} Gunn, J.E. and Peterson, B.A. 1965, \apj, 142, 1633
\bibitem{Holt89} Holtzman, J.A. 1989, \apjs, 71, 1
\bibitem{hu94} Hurley, K., \etal\ 1994, Nature, 372, 652.
\bibitem{} Kane, G.L. \& Wells, D. 1996, Phys. Rev. Letters, 76, 4458
\bibitem{Klyp95} Klypin, A., {\it et al.} 1995, \apj, 444, 1.
\bibitem{knif96} Kniffen, D.A., \etal\ 1996, A\&AS, 120, 615.
\bibitem{la95} Lanzetta, K.M., Wolfe, A.M., \& Turnshek, D.A. 1995,
	\apj, 440, 435
\bibitem{Lidd96} Liddle, A.R., {\it et al.}, 1996, MNRAS, 281, 531
\bibitem{li96} Lilly, S.J., Le Fevre, O., Hammer, F., \& Crampton, D.
	1996, \apj, 460, L1
\bibitem{MaBe94} Ma, C.-P. \& Bertschinger, E. 1994, \apj, 434, L5
\bibitem{mac} MacMinn, D \& Primack, J. 1996, Space Sci. Rev., 75, 413.
\bibitem{ma92} Madau, P. 1992, \apj, 389, L1
\bibitem{ma95} Madau, P. 1995, \apj, 441, 18
\bibitem{ma94} Madau, P. \&  Meiksin, A. 1994, \apj, 433, L53
\bibitem{mad96} Madau, P. 1996, in {\it Star Formation Near and Far}, 
  AIP Symp. Proc. (New York: Amer. Inst. Phys.), in press
\bibitem{mf96} Madau, P. \etal 1996, MNRAS 283, 1388
\bibitem{mp96} Madau, P. \& Phinney, E.S. 1996, \apj 456, 124
\bibitem{ma96} Madau, P. \& Shull, J.M. 1996, \apj, 457, 551
\bibitem{mr89} Martin, C. \& Bowyer, S. 1989, \apj 338, 677
\bibitem{mt90} Mattila, K. 1990, in IAU Symp. Vol. 139, The Galactic and
   Extragalactic Background Radiation, D. S. Bowyer \& C. Leinert (Dordrecht:
   Kluwer) 257
\bibitem{me93} Meiksin, A. \& Madau, P. 1993, \apj, 412, 34
\bibitem{mi92} Miralda-Escud\'{e}, J. \& Ostriker, J.P. 1992, \apj,
	392, 15
\bibitem{mi90} Miralda-Escud\'{e}, J. \& Ostriker, J.P. 1990, \apj,
	350, 1
\bibitem{pa80} Paresce, F., McKee, C., \& Bowyer, S. 1980, \apj, 240, 29
\bibitem{pe95} Pei, Y.C. \& Fall, S.M. 1995, \apj, 454, 69
\bibitem{pe94} Pettini M., Smith, L.J., Hunstead, R.W., \& King, D.L.
	1994, \apj, 426, 79
\bibitem{Pogo93} Pogosyan, D.Yu. \& Starobinsky, A.A. 1993, MNRAS, 265, 507 
\bibitem{} Pogosyan, D.Yu. \& Starobinsky, A.A. 1995, \apj\ 447, 465
\bibitem{Prim95} Primack,, J. {\it et al.} 1995, Phys. Rev. Lett. 74, 2160
\bibitem{Pro96} Protheroe, R. and Biermann, P.L. 1996, Astropart. Phys., 6, 45
\bibitem{Ruda89} Rudaz, S. 1989, Phys. Rev., D39, 3549
\bibitem{Ruda91} Rudaz, S. \& Stecker, F.W. 1991, \apj, 368, 40
\bibitem{Ruda88} Rudaz, S. \& Stecker 1988, F.W. \apj, 325, 16
\bibitem{sa94} Salamon, M.H., Stecker, F.W., \& De Jager, O.C. 1994,
	\apj, 415, L1
\bibitem{sa79} Savage, B.D. \& Mathis, J.S. 1979, \araa, 17, 73
\bibitem{Scha89} Schaefer, R.K., Shafi, Q. \& Stecker, F.W. 1989 \apj, 347, 575
\bibitem{sc91} Schneider, D.P., Schmidt, M., \& Gunn, J.E. 1991,
	\aj, 102, 837
\bibitem{Shaf84} Shafi, Q. \& Stecker, F.W. 1984, Phys. Rev. Lett. 53, 1292
\bibitem{Stec88} Stecker, F.W. 1988, Phys. Lett., B201, 529
\bibitem{ste96} Stecker, F.W. 1996 {\it Unveiling the Cosmic Infrared 
        Background (AIP Conf. Proc. 348)}, (Amer. Inst. of Physics, New York),
        p. 181.
\bibitem{stec96} Stecker, F.W. \& de Jager, O.C. 1996, Space Sci. Rev., 
        75, 413.
\bibitem{st97} Stecker, F.W. \& De Jager, O.C. 1997 \apj, 476, 712.
\bibitem{st92} Stecker, F.W., De Jager, O.C, \& Salamon, M.H. 1992,
	\apj, 390, L49
\bibitem{sds96} Stecker, F.W., De Jager, O.C, \& Salamon, M.H. 1996,
        \apj, 473, L75
\bibitem{st96} Stecker, F.W. \& Salamon, M.H. 1996, \apj, 464, 600
\bibitem{Stek89} Stecker, F.W. \& Tylka, A.J. 1989, \apj, 343, 169
\bibitem{st87} Steidel, C.C. \& Sargent, W.L.W. 1987, \apj, 318, L11
\bibitem{Th96} Thompson, D., \etal\ 1996, \apjs, 107, 227
\bibitem{vo95} Vogel, S.N., Weymann, R.J., Rauch, M. \& Hamilton, T. 1995,
   \apj\ 441, 162
\bibitem{wh92} Whittet, D.C.B. 1992, {\it Dust in the Galactic Environment},
	(Insitute of Physics Publishing, Bristol)
\bibitem{wo94} Worthey, G. 1994, \apjs, 95, 107

\end{thebibliography}
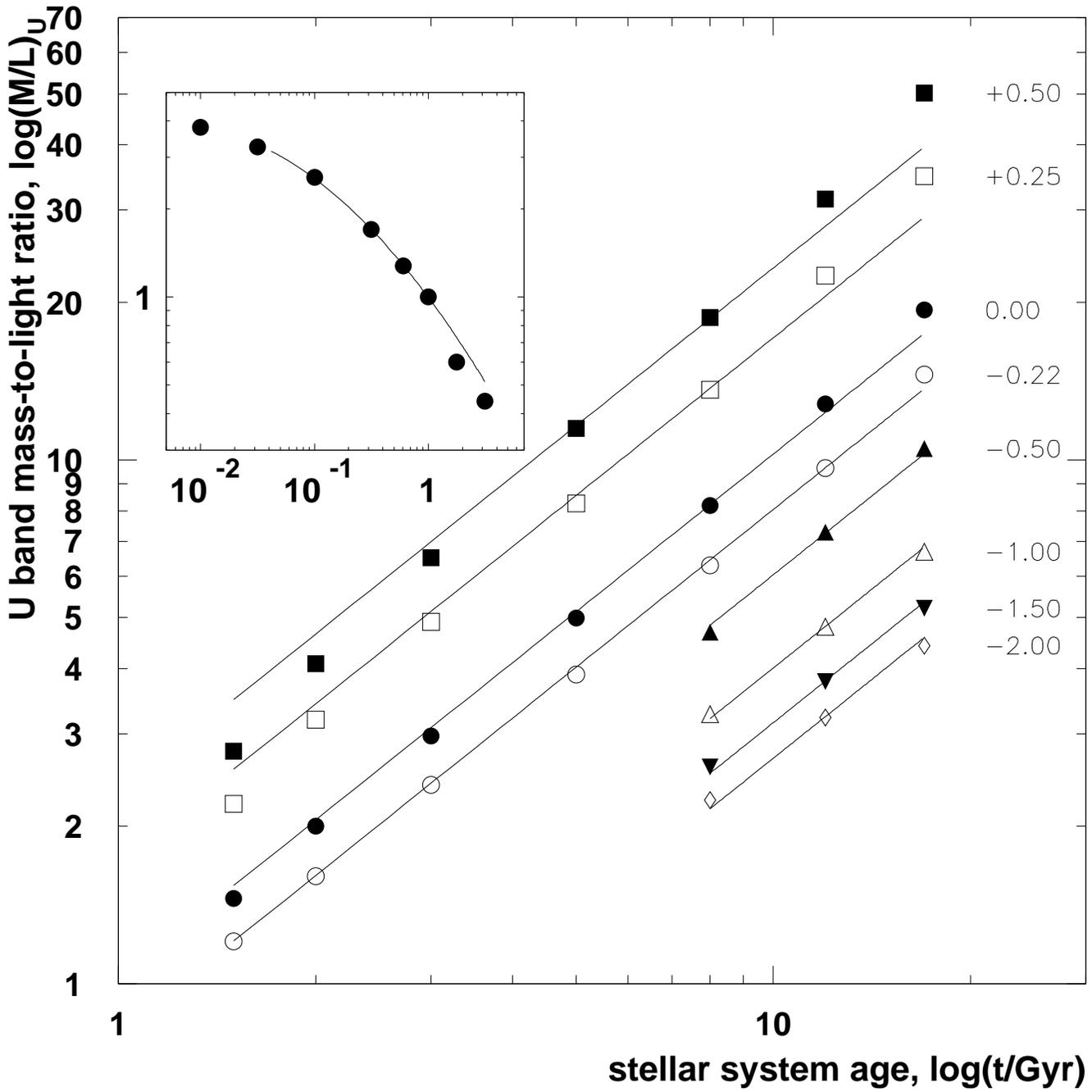
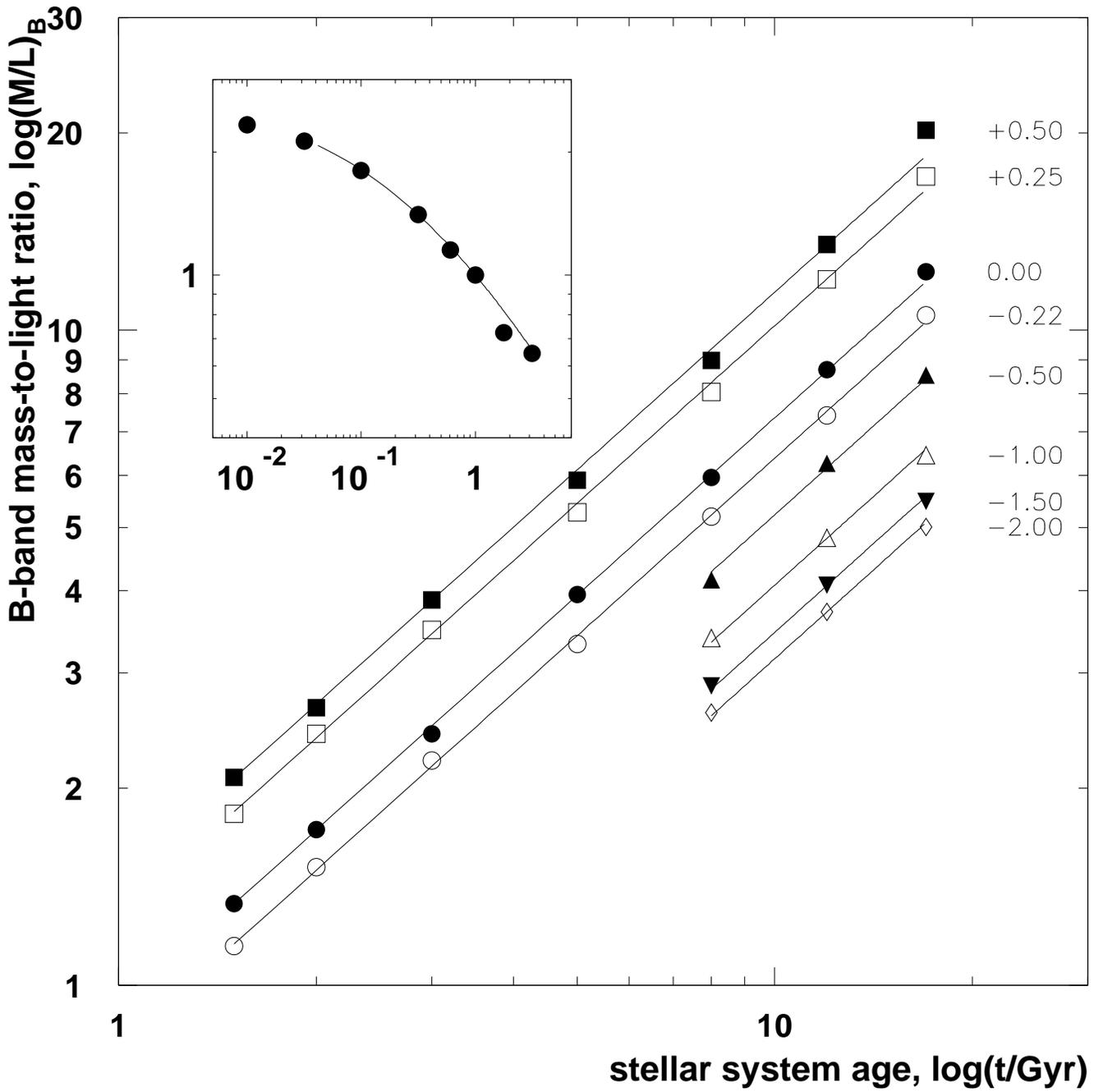
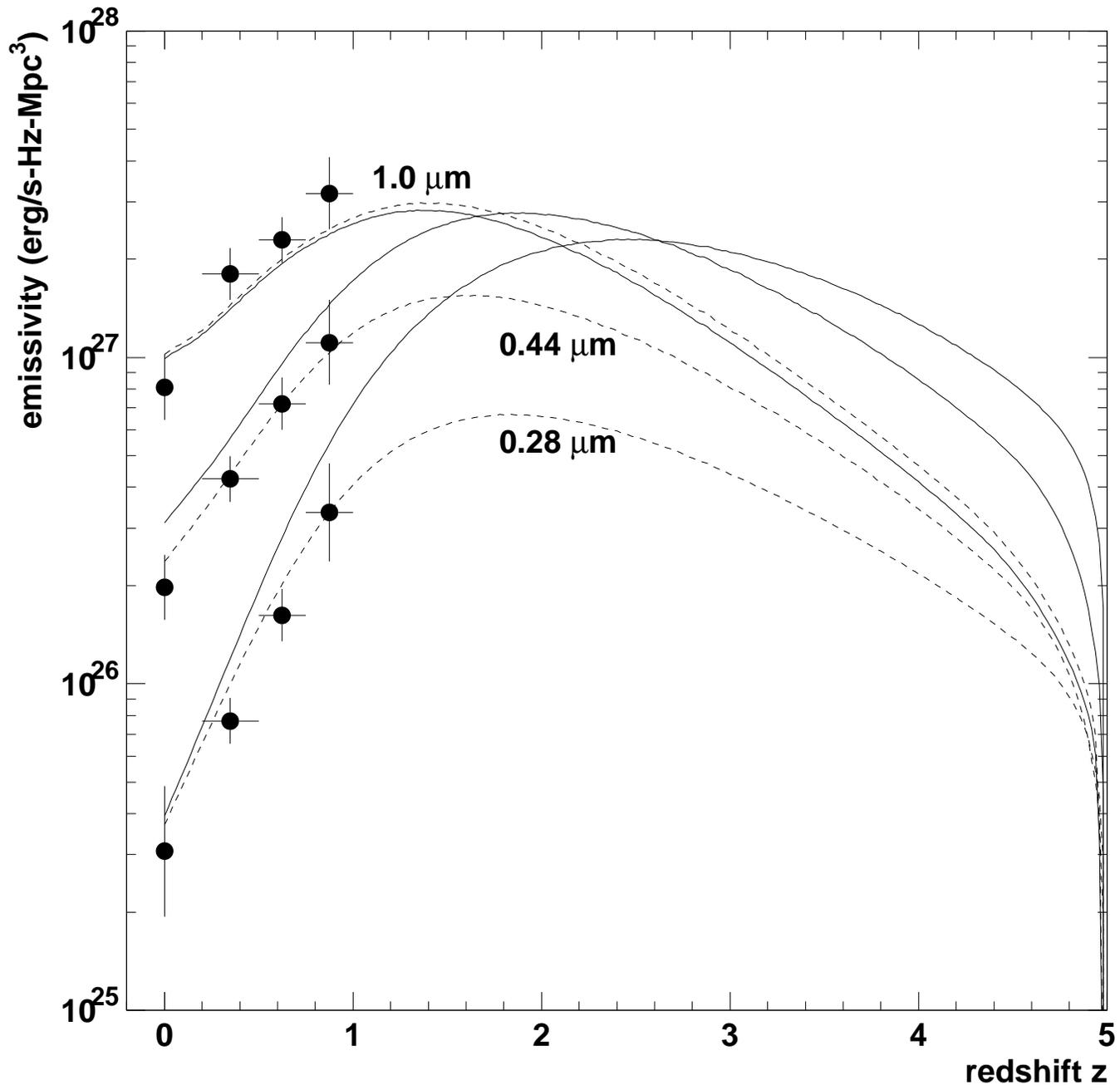
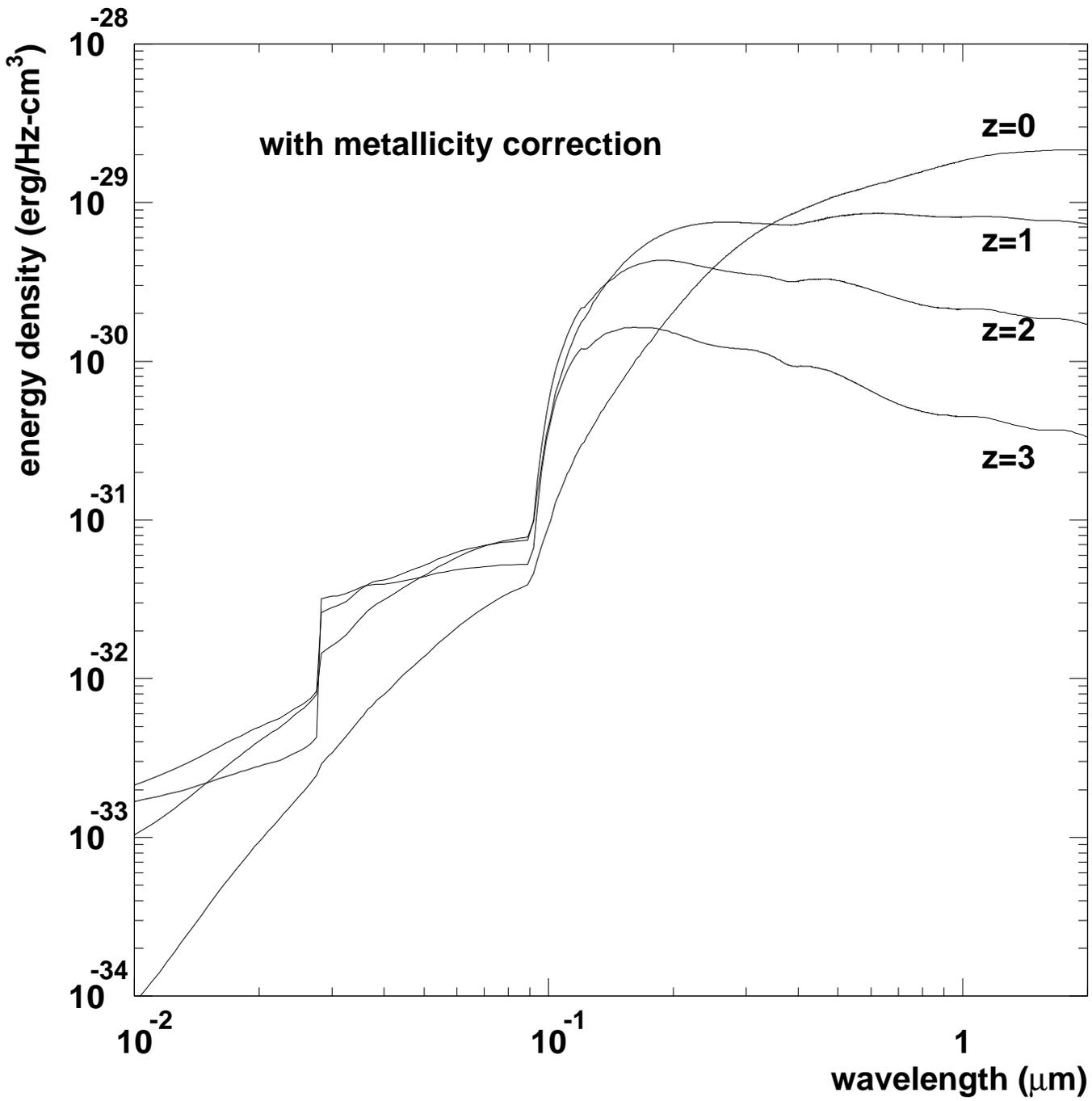
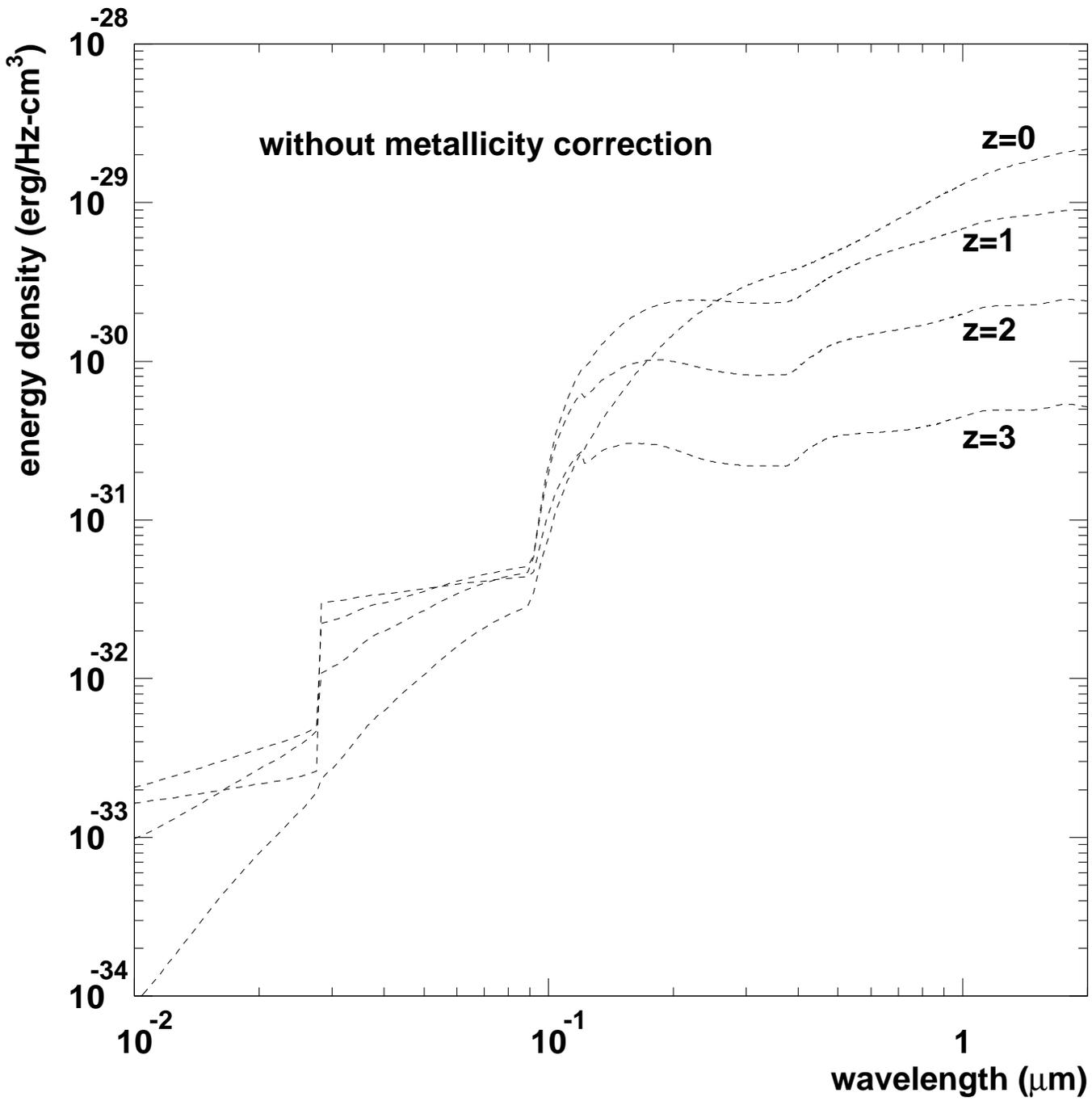
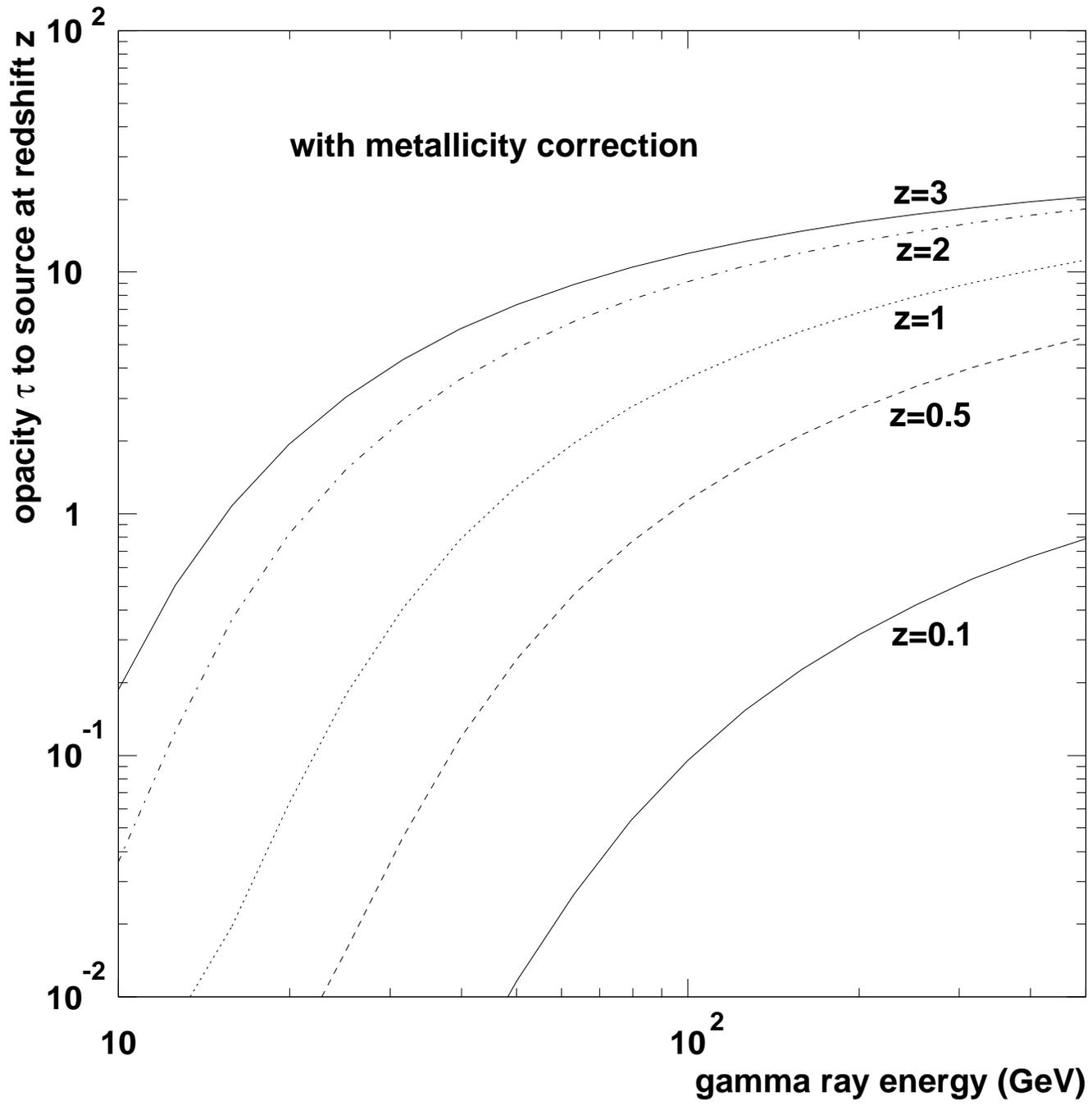
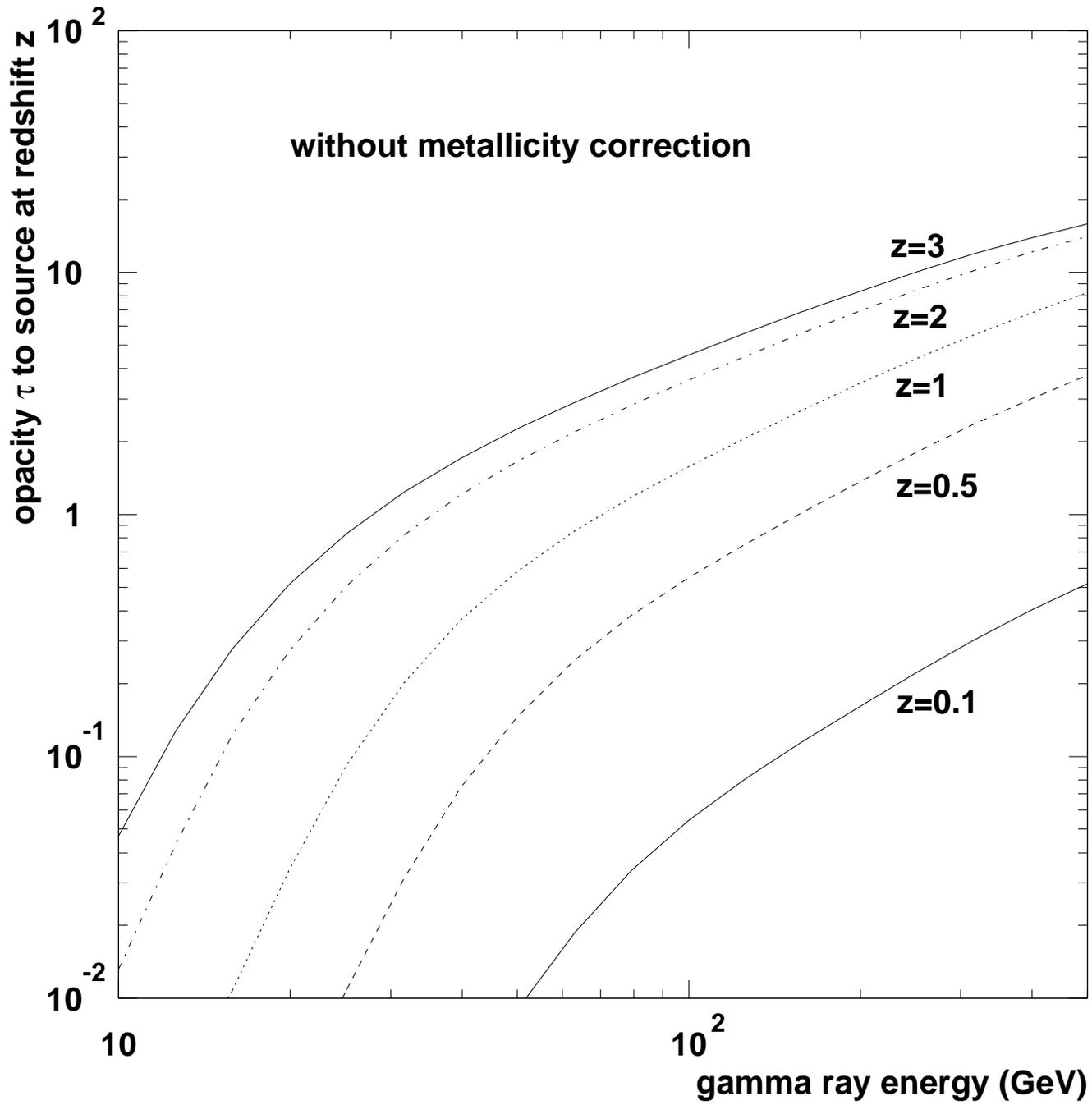
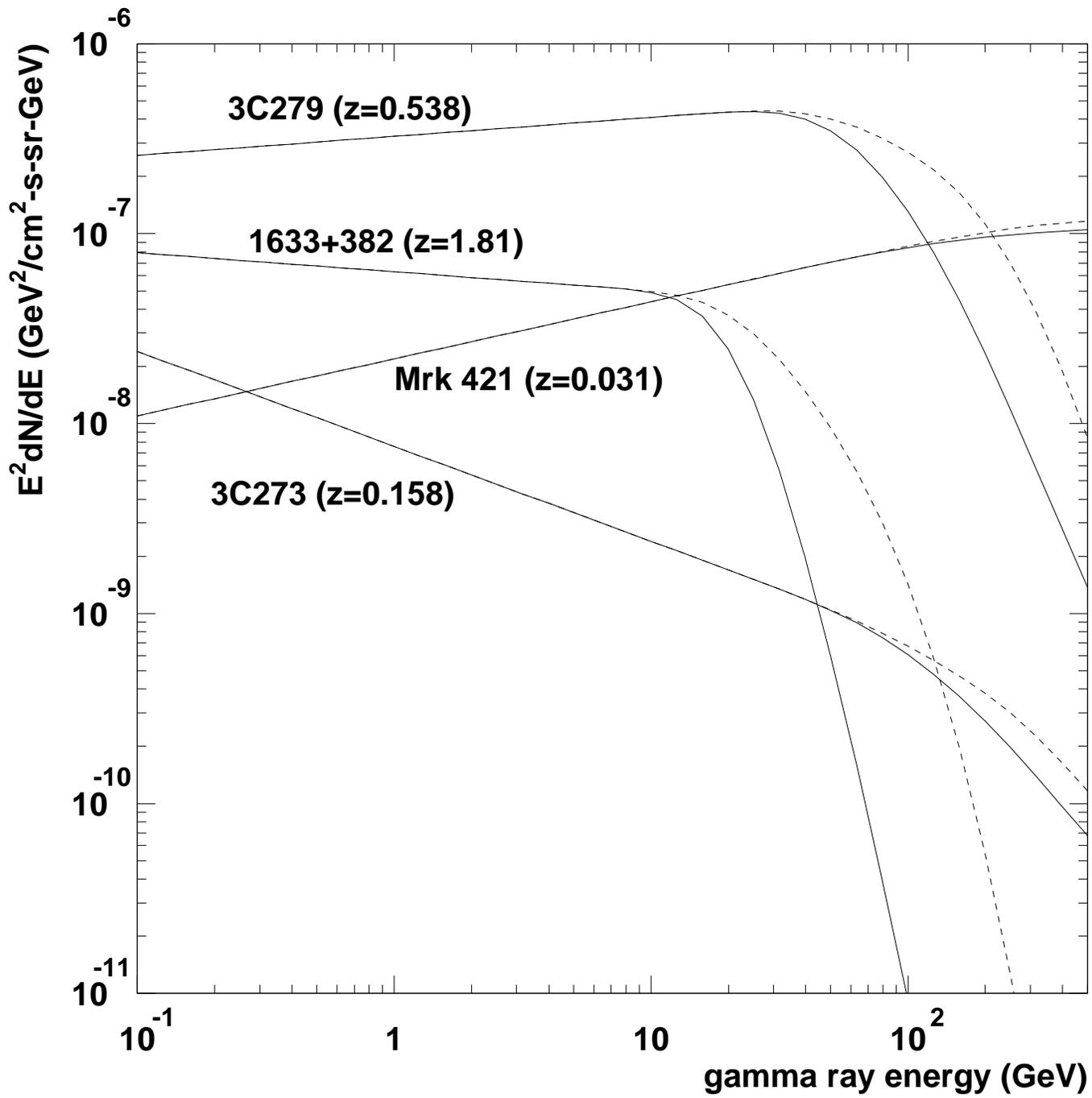
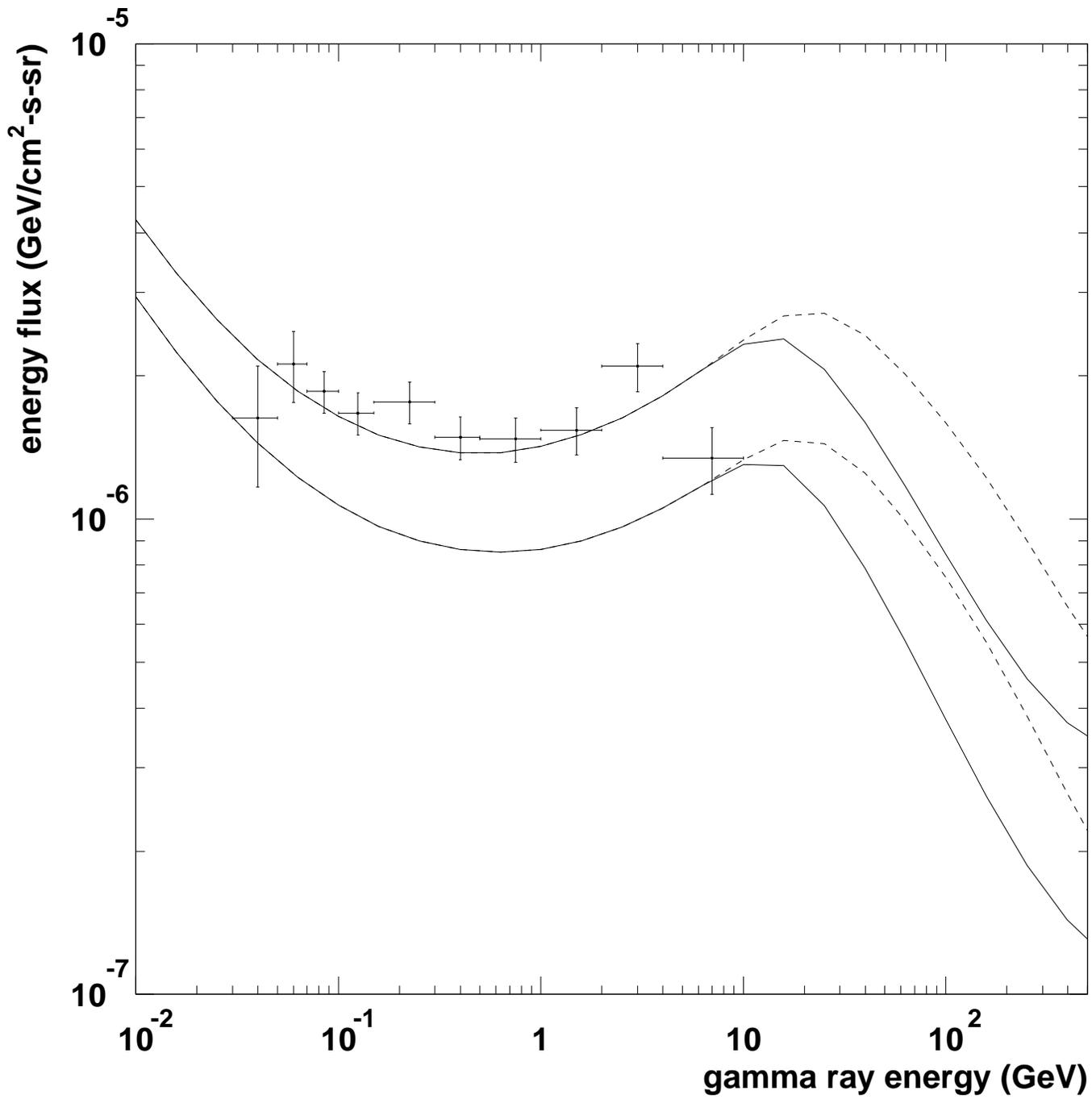
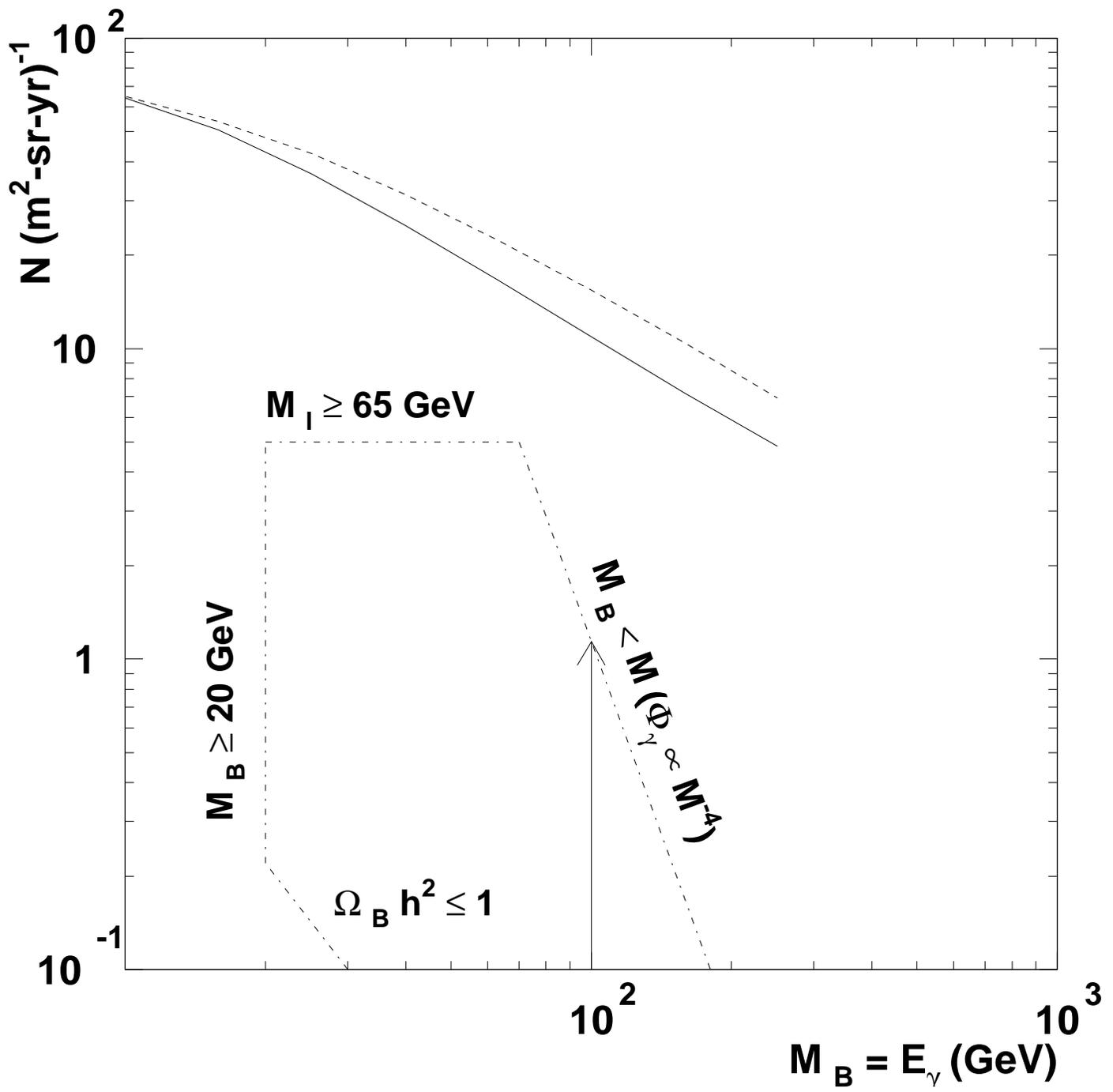
\end{document}